# Deciphering Carrier Dynamics in Polycarbonate Following Excitation with Ultrashort Laser Pulses

George D. Tsibidis[1,a)], Matina Vlahou [1,2], Stella Maragkaki [1], Ioannis Konidakis [1] and Emmanuel Stratakis [1,3,b)]

[1]*Institute of Electronic Structure and Laser (IESL), Foundation for Research and Technology (FORTH), Vassilika Vouton, 70013, Heraklion, Crete, Greece*

[2]*Department of Materials Science and Technology, University of Crete, 71003, Heraklion, Greece*

[3]*Department of Physics, University of Crete, 71003, Heraklion, Greece*

[a,b)] Authors to whom correspondence should be addressed: tsibidis@iesl.forth.gr; stratak@iesl.forth.gr

**ABSTRACT**

Polymers exposed to ultrashort pulsed lasers (UPL) experience a range of physical and chemical changes that play a key role in applications ranging from material processing to advanced photonics, and biomedicine. To elucidate the interaction of UPL with polymeric materials, ultrafast phenomena such as carrier dynamics, recombination and relaxation are investigated assuming polycarbonate (PC) as a test material exposed to laser pulses of moderate energies. A theoretical model developed for dielectric materials is extended to describe the, previously, unexplored excitation and carrier dynamics for PC while femtosecond Transient Absorption Spectroscopy is used to elucidate the evolution of the material's response and ultrafast dynamics. Interpreting the experimental measurements using the theoretical model suggests the existence of an energy level that facilitates the formation of self-trapped exciton metastates between the conduction and valence bands (approximately 2.4-2.8 eV below the conduction band). It also predicts the electron-plasma lifetime (around 110-150 fs), the recombination time (about 34 ps), and the non-linear part of the refractive index due to Kerr effect (with $n_2$ values ranging from 1.1-1.5×10$^{-16}$ cm$^2$/W). Furthermore, the dominant character of multi-photon assisted ionisation is emphasised while the optical breakdown threshold is also calculated and found to be equal to 2.55 ×10$^{12}$ W/cm$^2$. The results are expected to support future efforts aimed at elucidating how intense ultrashort laser pulses interact with polymeric materials which is crucial for optimizing the manufacturing processes of these materials for potential applications.

## 1. Introduction

The employment of femtosecond (fs) pulsed laser sources for material processing offers a highly precise method with significant potential to create a wide range of topographies in various types of materials for diverse technological applications [1-4]. Recent interest has grown considerably in surface functionalization of polymers, motivated by their key applications in areas such as tissue engineering and drug delivery, owing to their excellent biocompatibility [5]. It is known, that in order to empower a controlled fabrication of surface patterns on solids, it is crucial to understand the fundamental physical processes that govern the interaction between laser and the irradiated material. Despite the numerous studies on the response of various types of materials (e.g., metals, dielectrics, semiconductors, ceramics) [6-9] to irradiation with ultrashort pulsed lasers, a comprehensive framework addressing the underlying physical mechanisms in polymer patterning remains elusive. A major challenge in understanding the topography fabrication process is that, unlike other materials, exposure of polymers to high-intensity laser pulses can lead not only to structural photo-induced changes but also to alterations in the material itself (such as photo-chemical degradation, chain rearrangements, bond breaking, or polymerization) [10].

To address the above challenges, a proof-of-concept approach is pursued to explore the ultrafast dynamics that occur in Polycarbonate (PC) exposed to laser pulses of moderate energies. These energies have been appropriately selected to ensure they are sufficient to excite the material, while avoiding interference from complex phenomena that could alter the polymer's characteristics or induce damage, thereby allowing for a direct elucidation of the underlying processes. The focus of the investigation is on modeling the dynamics of excited electrons while Transient Absorption Spectroscopy (TAS) is used to evaluate ultrafast temporal changes of the excited states in polymers. The advantage of selecting TAS over traditional pump-probe techniques [11] is that it allows the study of photo-induced absorption features that are correlated with free charge carrier dynamics and recombination processes occurring in picosecond regime. Such phenomena are known to strongly determine the optical and electronic characteristics of materials used in a wide range of applications that include optical materials [12], energy harvesting applications [13, 14], and photocatalysis [15]. The operation principle of TAS relies on a light source that is used to photoexcite electrons while the corresponding dynamics of the relaxation processes are monitored in terms of optical absorption within various time delays [12-15]. Previous research has focused on assessing absorptivity under different excitation conditions and modelling the evolution of electron-hole concentrations. This was achieved through the development of a kinetic model that incorporates a single shallow trap state, along with associated trapping rate constants [16, 17]. Nevertheless, to the best of our knowledge, such an analysis is ineffective at directly including the impact of laser energies on the excitation process, the types of ionization mechanisms (such as multiphoton, tunnelling, and impact ionization) or relaxation processes.

The objective of the present study is to address the above issues and provide a consistent exploration of the underlying physical mechanisms that account for the ionisation process. Therefore, in this work, we aim to provide a detailed analysis of the laser interaction with dielectric materials that accounts for the material ionization and the electron-plasma relaxation considering a combined theoretical/experimental approach. To validate the methodology, an established model and TAS data—both employed to describe electron excitation and dynamics—are initially applied to a well-characterized material, fused silica, before being used to discuss the previously unexplored ionization processes in PC.

## 2. Theoretical Model

In the present work and due to the high transparency of polymeric material, a theoretical framework developed for *dielectric* materials is used to describe the free electron generation and evolution of the excited electron density $N_e$ [18, 19] in PC upon excitation with ultrashort pulsed lasers (see Eq.1 [18, 20-22])

$$\begin{aligned}
\frac{dN_e}{dt} &= \frac{N_V - N_e}{N_V}\left(W_{PI}^{(1)} + N_e \mathbb{A}^{(1)}\right) + \frac{N_{STE}}{N_V}\left(W_{PI}^{(2)} + N_e \mathbb{A}^{(2)}\right) - \frac{N_e}{\tau_{tr}} \\
\frac{dN_{STE}}{dt} &= \frac{N_e}{\tau_{tr}} - \frac{N_{STE}}{N_V}\left(W_{PI}^{(2)} + N_e \mathbb{A}^{(2)}\right) - \frac{N_{STE}}{\tau_{rec}}
\end{aligned} \quad (1)$$

where $N_V$~$2.5\times10^{21}$ cm$^{-3}$ [23] corresponds to the valence electron density for PC. In the above framework, photoexcitation ($W_{PI}^{(i)}$ stands for the photoionization rate) and impact ionization processes ($\mathbb{A}^{(i)}$ stands for the avalanche ionisation rate) allow excitation of electrons from the valence (VB) to the conduction band (CB), where the energy band gap of PC is equal to $E_G^{(1)}$ =4.3 eV [24]. The index $i$ corresponds to describe excitation of electrons from the VB to CB ($i$=1) or from the energy level where self-trapped exciton (STE) states reside are excited to CB ($i$=2) [20, 21]. Although simpler models can be employed under the assumption of the absence of STE states generation [25, 26], the fact that exciton states are also linked to defect formation—a common characteristic in polymers (i.e. they have a higher concentration of impurities and structural defects)—implies that overlooking STE states could lead to an oversimplified description of the ultrafast processes.

In Eq.1, $\tau_{tr}$ is the characteristic time of free electron decay (i.e. including recombination and trapping in metastable STE); $\tau_{tr}$ varies between 90 fs and some picoseconds (ps) [27]. Moreover, $\tau_{rec}$ standing for the exciton decay time (~some ps) [22] and $E_G^{(2)}$ representing the band gap between the STE level and CB, are taken as parameters to be determined. An attenuation of the local laser intensity $I$ due to the photoionization and inverse bremsstrahlung absorption along the depth $z$ of the material is considered and described by Eqs.2

$$\begin{aligned}
\frac{dI}{dz} &= -N_{ph}^{(1)} \hbar\omega_L \frac{N_V - N_e}{N_V} W_{PI}^{(1)} - N_{ph}^{(2)} \hbar\omega_L \frac{N_{STE}}{N_V} W_{PI}^{(2)} - \alpha_{FCA}(N_e) I \\
I &= (1-R)\frac{2\sqrt{\ln 2}}{\sqrt{\pi}\tau_p} F e^{-4\ln 2\left(\frac{t-3\tau_p}{\tau_p}\right)^2}
\end{aligned} \quad (2)$$

where $N_{ph}^{(1)}$ (or $N_{ph}^{(2)}$) stands for the minimum number of photons required to be absorbed by an electron located in the valence band (or STE energy level) to be excited to the conduction band, $\omega_L$ corresponds to the laser frequency assuming excitation of the solid at $\lambda$, $F$ is the peak fluence and $1-R$ stands for the part of the energy transmitted through the surface of PC. The shift $3\tau_p$ in the second expression in Eq.2 is used to ensure that energy deposition ceases when the pulse is switched off at time $t$=$6\tau_p$. The reflectivity $R$ is derived from the dielectric parameter [18, 28]

$$\varepsilon = 1 + (\varepsilon_{un} - 1)\left(1 - \frac{N_e}{N_V}\right) \frac{e^2 N_e}{m_r m_e \varepsilon_0 \omega_L^2} \frac{1}{\left(1 + i\frac{1}{\omega_L \tau_c}\right)} \quad (3)$$

In Eq.3, $\varepsilon_{un}$ corresponds to the dielectric parameter of the unexcited material [29], $m_e$ is the electron mass, $e$ is the electron charge, $m_r = 0.5 - 1$ ($m_r$ denotes the optical effective mass of the carriers) [27], $\varepsilon_0$ is the vacuum permittivity. Various values for $\tau_c$, the electron collision time, have been used in previous reports (see Ref. [18] and references therein), however, here, $\tau_c$ is taken equal to $\tau_c = 1.1$ fs. A contribution due to Kerr effect is also introduced in the dielectric parameter; thus, an effective value $\varepsilon_{eff} = \varepsilon + 2n_0 n_2 I + (n_2 I)^2$ is considered where $n_2$ stands for the nonlinear part of the refractive index of PC while $n_0 = \sqrt{\varepsilon_{un}} = 1.5644$ [29]. Whereas $n_2$~$10^{-16}$ cm$^2$/W for fused silica [18], $n_2$ for PC is taken as a parameter to be determined. Finally, $\alpha_{FCA}(N_e)$ in Eq.2 is the free carrier absorption coefficient.

## 3. Experimental Protocol

In the present work, TAS measurements were performed on a Newport transient absorption spectrometer, equipped with a source pulsed laser beam generated from an Yb:KGW-based laser system, emitting at a laser wavelength $\lambda$=1026 nm with a pulse duration

$\tau_p = 170$ fs and 1 kHz repetition rate [12-15]. The 1026 nm fundamental beam was split (Figure 1), so that the probe beam component (10% of the source) passes through a delay line and routed on a YAG crystal, which generates a supercontinuum white light of 550-920 nm. The other part of the incident beam (90% of the source) was used as the pump beam for sample excitation. The energy of the pump beam was controlled by a variable reflective neutral density filter inside the TAS instrument. The probe light was coupled through an optical fiber to a multichannel detector and monitored as a function of wavelength.

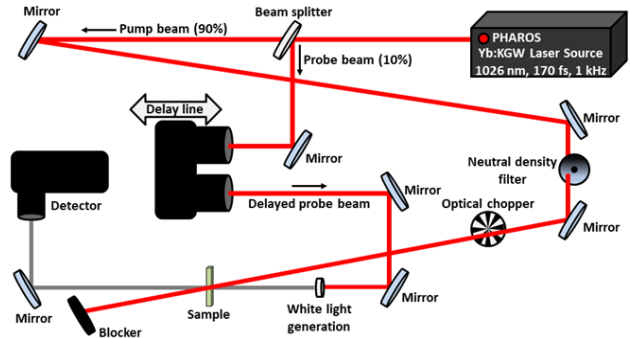

**Figure 1:** Schematic representation of TAS experimental setup [12-15].

In a standard TAS pump-probe experiment, the sample is excited by the pump beam while the resulting decay dynamics of the sample's relative optical density are measured as a function of wavelength at different time delays following photo-excitation. The excitation of the sample by the laser pump beam induces changes in the sample's spectrum. These spectra changes are monitored at many wavelengths simultaneously by a white light supercontinuum probe as sensed by the combination of a spectrograph and a photodiode array. To reduce the effect of laser fluctuations, the sample's absorption is measured first without the pump, with the optical chopper blocking the beam, and then the pump unblocked using the subsequent pulse of the laser. The time between pump and probe is adjusted using a delay stage. For the probe pulses, a supercontinuum white light in the 1.12 eV to 2.34 eV range of photon energies was generated using a YAG crystal. The time resolution was 250 fs. The peak fluence $F$ was calculated by the relation $F = 2E_P/\pi r^2$, where the pulse energy $E_P$ was derived from the pump power measured with a power meter. The spot size, measured using a beam profiler, was found to have a 1/e² beam diameter of 2$r$=600 μm.

## 4. Results and Discussion

Experiments were performed in transmission geometry mode in selected time domains (0, 0.2, 0.3, 0.5, 0.7, 1) ps. The measured difference in optical density ($\Delta OD$) has been set as $\Delta OD$ = log(blocked/unblocked) = log(blocked) - log(unblocked) = OD$_{probe}$ - OD$_{pump+probe}$ [12-15]. Based on this definition, a negative signal corresponds to ground state photobleaching, or a depletion of the population of the ground state, indicating that there are fewer photoreceptors in the ground state to absorb the probe light. By contrast, when the signal is positive, less light is detected after the sample is excited, indicating photoabsorption in the excited state. In this study, excitation pump fluences within the range of 14-26 mJ/cm² were employed, while measurements were performed under ambient lab conditions.

The difference in optical density over wavelength at different time delays are shown in Figs. 2-3, corresponding to fused silica and polycarbonate, respectively. In both silica and polycarbonate cases the obtained $\Delta OD$ signal decays with time (time delay values are: 0, 0.2, 0.3, 0.5, 0.7 1) ps, as expected due to the resulting charge carrier relaxation phenomena. Meanwhile, for both samples (Figs. 2-3) the $\Delta OD$ levels increase upon employing higher pump fluence, i.e. higher population of excited carriers.

For fused silica (Figure 2), the negative $\Delta OD$ indicates a photobleaching effect. In this scenario, the probe pulse excites carriers from the ground state to higher energy states, and the resulting decay over time reflects the relaxation of the excited carriers back to lower energy states. Conversely, for polycarbonate (Figure 3), the positive $\Delta OD$ suggests that additional carrier excitation occurs from higher energy levels before photoionization takes place. The observed optical density evolution is attributed to the relaxation of these carriers to lower energy states [12-15]. It is important to note that the analysis in this study was performed using the wavelengths where $\Delta OD$ showed the greatest absolute magnitude (i.e., 1011 nm for fused silica and 785 nm for polycarbonate), enabling the approach to draw definitive conclusions.

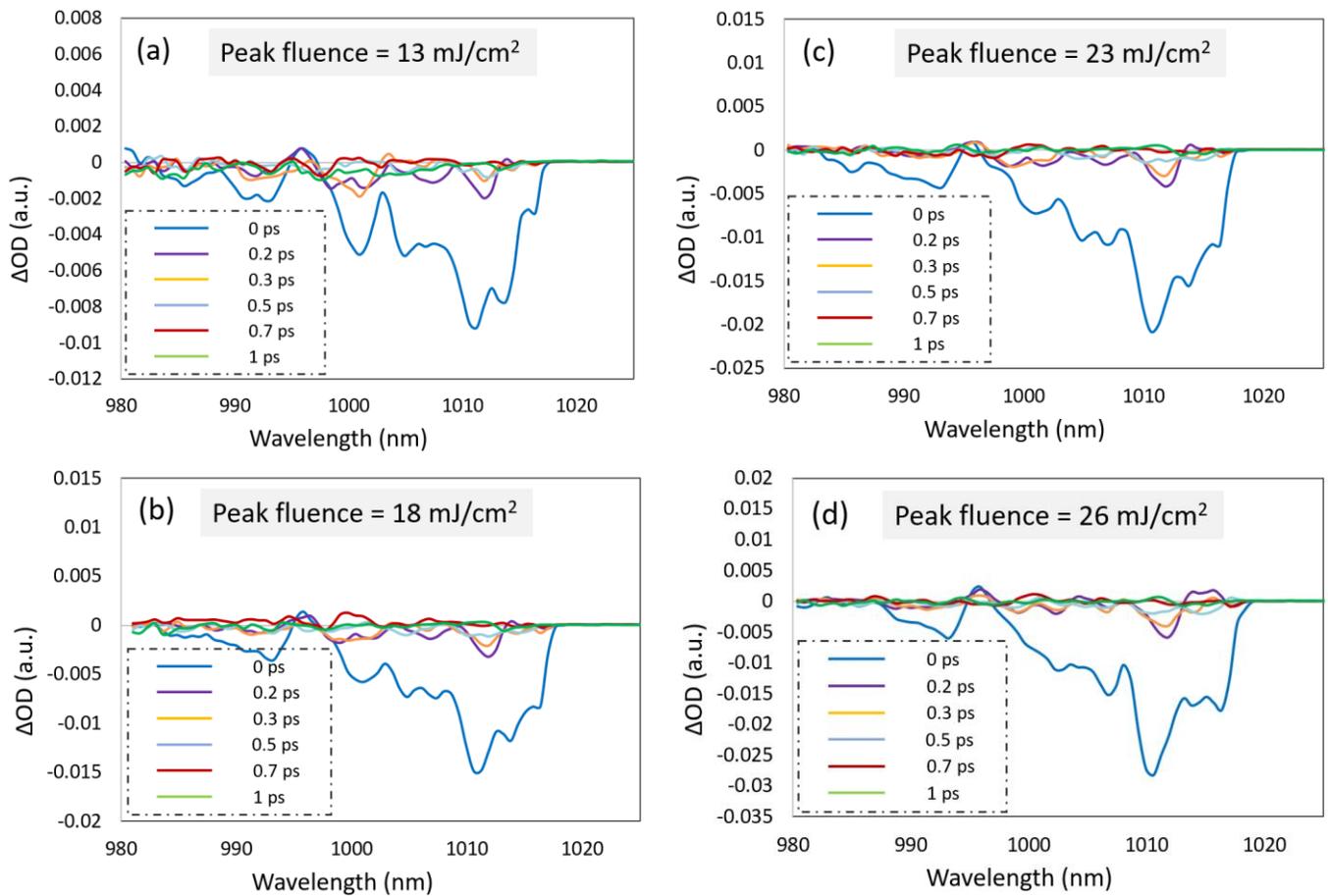

**Figure 2:** Difference in optical density as a function of wavelength at time delays (0, 0.2, 0.3, 0.5, 0.7, 1) ps after photo-excitation of fused silica at four different laser fluence values: (a) $F$=13 mJ/cm$^2$, (b) $F$=18 mJ/cm$^2$, (c) $F$=23 mJ/cm$^2$, (d) $F$=26 mJ/cm$^2$.

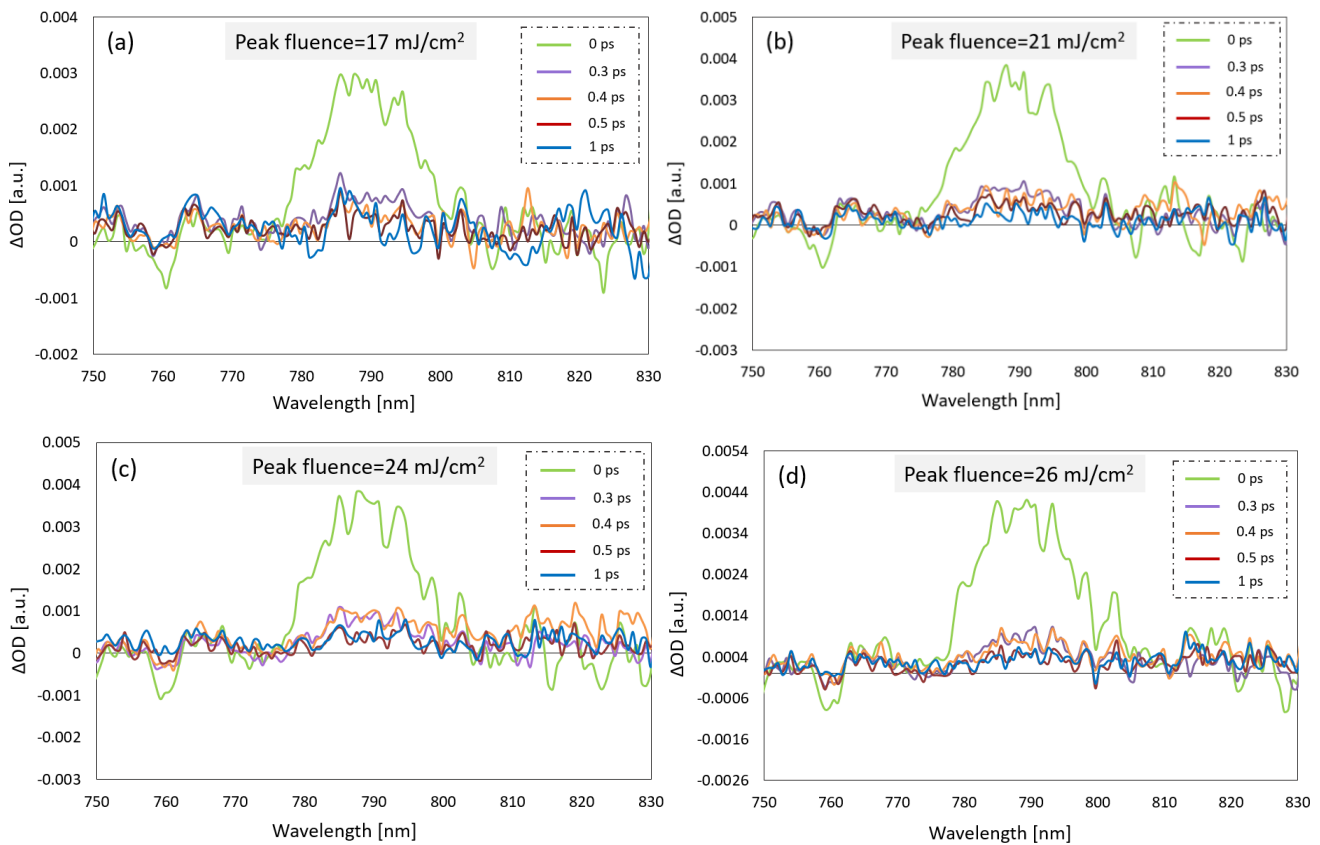

**Figure 3:** Difference in optical density as a function of wavelength at time delays [0, 0.2, 0.3, 0.5, 0.7, 1] ps after photo-excitation of Polycarbonate at four different laser fluence values: (a) $F$=17 mJ/cm$^2$, (b) $F$=21 mJ/cm$^2$, (c) $F$=24 mJ/cm$^2$, (d) $F$=26 mJ/cm$^2$.

Before applying the theoretical model to PC, the validity of the approach for correlating experimental data from TAS with simulation results was first evaluated using a well-characterized dielectric material, fused silica. For this material, the theoretical framework and the values of the free parameters have been verified to produce accurate results for electron dynamics [16, 18-20]. TAS experiments were performed for a 1 mm thick film of fused silica and optical density values were obtained at time delays $\Delta t$ for fluences $F$=13 mJ/cm$^2$, 18 mJ/cm$^2$, 23 mJ/cm$^2$, 26 mJ/cm$^2$ (Figure 4a). Similar experiments were, also, conducted for a 0.5 mm thick PC film while optical density values were obtained for fluences $F$=17 mJ/cm$^2$, 21 mJ/cm$^2$, 24 mJ/cm$^2$, 26 mJ/cm$^2$ (Figure 4b). It is emphasised that optical density is negative in SiO$_2$, whereas it is positive for PC.

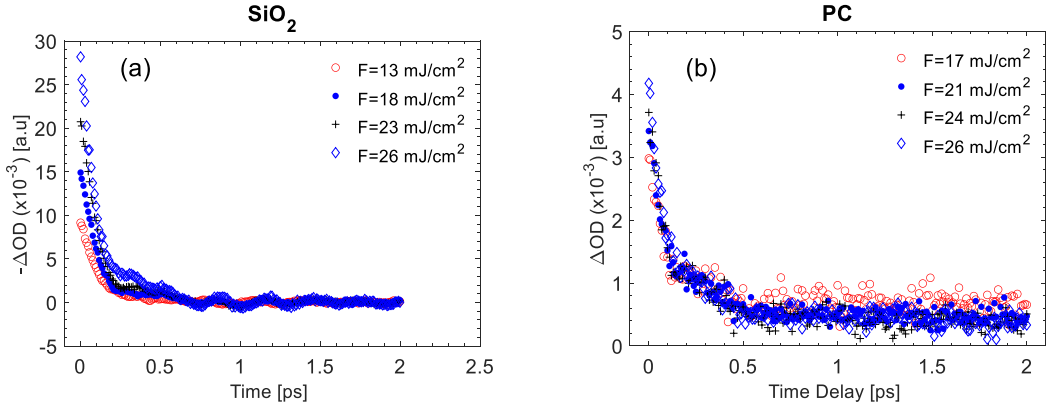

**Figure 4:** Evolution of optical density for (a) SiO$_2$ (at 1011 nm) and (b) PC (at 785 nm) photoexcited at 1026nm with various pump fluences for various fluences (illustrated at time delays $\Delta t$<2 ps).

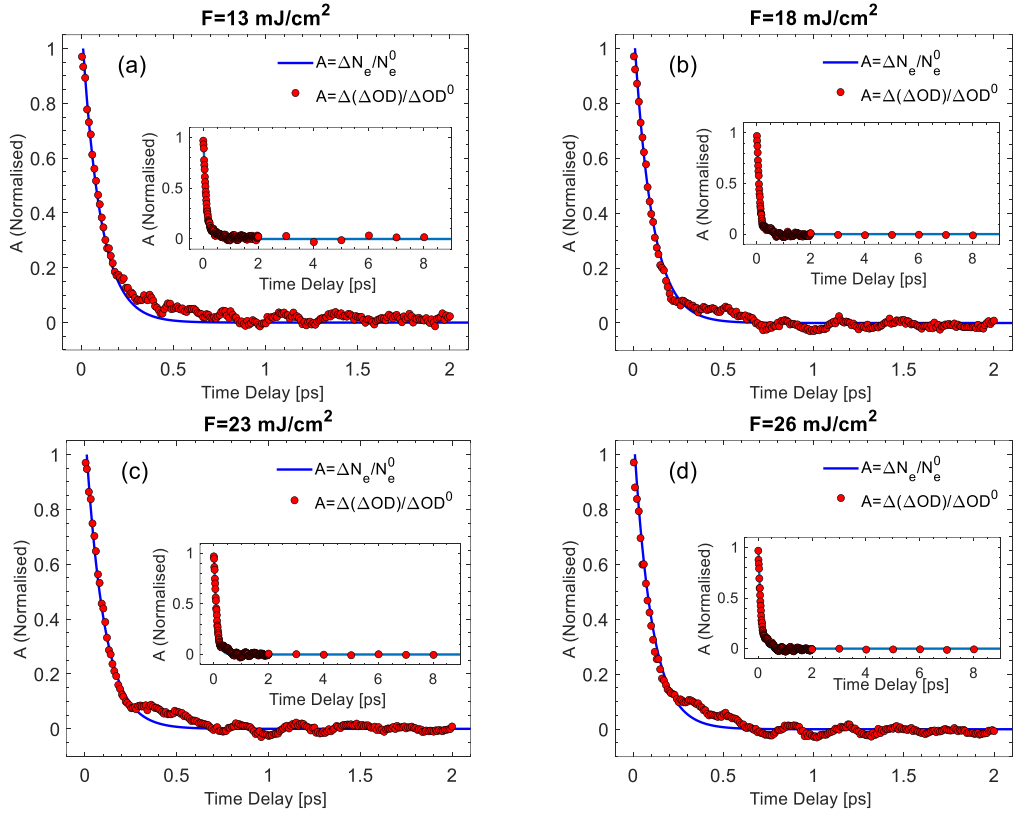

**Figure 5:** Evolution of optical (filled *red* circles) and carrier density (*blue* solid lines) variations in fused silica under photoexcitation at four different fluences (a) $F$=13 mJ/cm$^2$, (b) $F$=18 mJ/cm$^2$, (c) $F$=23 mJ/cm$^2$, (d) $F$=26 mJ/cm$^2$. The inset shows the evolution at longer time delays ($\Delta OD^0 \equiv \Delta OD^{(min)}$ and $N_e^0 \equiv N_e^{(min)}$).

In regard to the TAS analysis for *fused silica* (Figure 4a), $\Delta t$=0 represents the delay at which no further excitation is expected. The values correspond to results from the probe pulse measured when the pump pulse is switched off (at $t$=6$\tau_p$=1.1 ps). To correlate the variation in the optical density with the change in the density of the excited carriers (which is directly linked to the absorption levels achieved at different fluences), Eqs.1-3 are used. A fitting procedure is pursued to determine the parameters such as $\tau_{tr}$, $\tau_{rec}$, which define the relationship between the normalized variation of $\Delta OD$ ($\equiv (\Delta OD(t) - \Delta OD^{(min)})/\Delta OD^{(min)}$), and the carrier density ($\equiv N_e(t) - N_e^{(min)})/N_e^{(min)}$ (i.e. $(\Delta OD(t) - \Delta OD^{(min)})/\Delta OD^{(min)} \sim (N_e(t) - N_e^{(min)})/N_e^{(min)}$). The normalized variation of these quantities is illustrated in Figure 5 and results manifest two rapid components of characteristic value decrease, one sharp and the other slow. In principle, the fast component is strongly dependent on the photoionization conditions and the fluence (Figure 4a). The parameters predicted for fused silica are between $\tau_{tr}$=90 fs (for $F$=13 mJ/cm$^2$) and $\tau_{tr}$=135 fs ($F$=26 mJ/cm$^2$), while $\tau_{rec}$ = 34 ps (in practise, independent of the fluence); on the other hand, the values for $n_2, E_G^{(2)}$ were obtained from previous reports ($n_2 = 10^{-16}$ cm$^2$/W and $E_G^{(2)}$=6 eV) [18]. The estimated values for the timescales are in agreement with those reported in previous works [18, 22]. Fitting the TAS data to a bi-exponential function, a standard method to analyse observations from pump-probe experiments [17], produces a comparable range of values for the decay time of the fast component but not that of the slow component (as also emphasised in previous reports) [22]. The results indicate that the TAS technique and the theoretical model describe adequately the dynamics of the carrier system. It is emphasized that despite the negative $\Delta OD$ values (which are indicative of photobleaching effects and decreased absorption), the above theoretical framework, accurately predicts the relaxation of carrier density (see, also, Supplementary Material) following excitation with low-intensity femtosecond pulses; in particular, the decay of the $\Delta OD$ to the initial value agrees with the carrier density relaxation. The accuracy of a more complex model that includes terms accounting for bleaching is beyond the scope of the current study.

A similar analysis was performed for *PC* and the employment of Eqs.1-3 and a fitting procedure yielded $\tau_{tr}$=110-150 fs, $\tau_{rec}$ =34 ps, $n_2 = (1.1-1.5) \times 10^{-16}$ cm$^2$/W while $E_G^{(2)}$ =2.4-2.8 eV. Therefore, for the laser parameters used in this study, the decay components both for PC and fused silica appear to be characterised by comparable rates. The results shown in Figure 6 demonstrate that the dynamics is sufficiently described using Eqs. 1-3, despite some dispersion in the experimental data at low fluences. On the other hand, a thorough investigation of both the experimental and theoretical results require a closer examination of the differences between the characteristics of PC and fused silica and the irradiation conditions. Due to a lower band gap $E_G^{(1)}$, the generation of a significantly larger number of excited carriers are produced (see Supplementary Material). In particular, for fused silica, despite the low excitation level achieved at the fluences used in this study and the minimal absorption from the carriers (leading to a negative $\Delta OD$), the theoretical model effectively describes the relaxation process. Similar conclusions were also reached for the polymeric material where irradiation at similar fluences led to higher excitation level, higher absorption and positive $\Delta OD$.

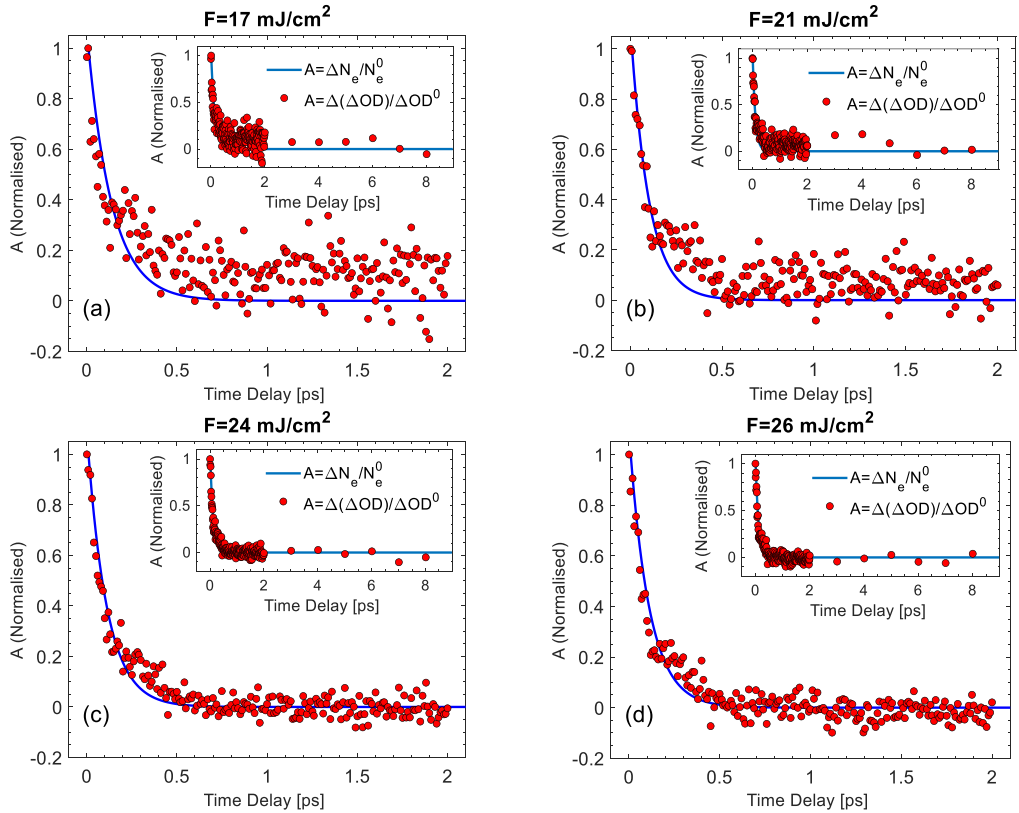

**Figure 6:** Relaxation of optical (filled *red* circles) and carrier density (*blue* solid lines) in PC under photoexcitation at four different fluences (a) $F$=17 mJ/cm$^2$, (b) $F$=21 mJ/cm$^2$, (c) $F$=24 mJ/cm$^2$, (d) $F$=26 mJ/cm$^2$ (filled *red* circles) and their corresponding decay fit (*blue* solid lines). The inset shows the evolution at longer times delays.

For the same laser parameters, the maximum number of the excited carrier densities are approximately ~$10^{10}$-$10^{12}$ times more in PC than for fused silica. Moreover, theoretical and experimental indicate that, unlike fused silica (at low fluences), a positive optical density is observed that is a characteristic of enhanced absorption (Figure 4).

One interesting observation in the TAS experiments (Figs. 4-6) is the emergence of coherent oscillations of the differential probe field in the pump-probe protocol for fused silica. Coherent phonon oscillations in semiconducting/dielectric materials, observed as a modulation in the amplitude of transient pump-probe spectra, have been reported and offered valuable insights into carrier-phonon interactions [16]. It has been observed that coherent oscillations emerge in pump-probe experiments across various materials [30, 31] and in a previous study, it was proposed that these oscillations are generated from electron-phonon coupling between the excited carriers and the longitudinal optical (LO) mode [16] resulting in oscillations in the transient signal. By contrast, excitation in PC (or other polymers) using the pump pulse may lead to more localized and incoherent vibrational modes, preventing the occurrence of similar effects (Figure. 4b); therefore, the oscillating character of the electron dynamics is not illustrated in Figure 6. Certainly, a more detailed theoretical framework would require to account for the processes that excite the vibrational modes and their interaction with the electron system (through the introduction of a revised thermal model [18] coupled with Density Functional Theory calculations [16]), however, it is beyond the scope of this work to focus on the impact of coherent oscillations on carrier dynamics.

One of the advantages of theoretical model is its ability to simulate excitation and carrier dynamics by examining the effects of photo-ionization and impact ionization processes. While these processes have been thoroughly explored in dielectric materials [20, 22, 32, 33], the specific regimes in which each type of photo-ionization (PI) dominates (i.e. multi-photon (MPI) or tunneling (TI)) have not yet been determined in polymeric materials. To achieve this, a systematic analysis is conducted to correlate the laser intensity with the type of photoionization that governs electron excitation in PC assuming irradiation at $\lambda$=1026 nm and $\tau_p$=170 fs; this methodology can, also, be applied to different wavelengths and pulse durations. As stated above, PI depends on the laser intensity, however, MPI and TI are efficient in different regimes of the intensity spectrum. In Figure 7a, the photoionization rate $W_{PI}$ is computed for (peak) intensities $I_0 = \frac{2\sqrt{\ln 2}}{\sqrt{\pi}\tau_p} F$ in the range between $10^9$ W/cm² and $10^{15}$ W/cm². To evaluate the regime where TI or MPI become more efficient in the PI process, the Keldysh parameter is also displayed (Figure 7b). The Keldysh parameter, $\gamma$ ~$1/(\sqrt{I_0}\lambda)$, indicates the regime in which each of the photo-ionization components, MPI and TI dominates [34]. It is evident that at large intensities, TI dominates ($\gamma \ll 1$) while at small intensities, MPI is the main contributor to PI ($\gamma \gg 1$). By contrast, an intermediate regime in which both TI and MPI coexist is illustrated in a region between the two regimes. According to simulation results shown in Figure 7a, for large intensities, the Keldysh approximation for MPI provides an underestimation of PI rates while for lower values of $I_0$ an overestimated TI is revealed. Thus, for the fluences used in this study, $F$=17 mJ/cm², 21 mJ/cm², 24 mJ/cm², 26 mJ/cm² that correspond to peak intensities equal to $0.9 \times 10^{11}$ W/cm², $1.16 \times 10^{11}$ W/cm², $1.32 \times 10^{11}$ W/cm², $1.43 \times 10^{11}$ W/cm², respectively, MPI accounts for the ionisation process.

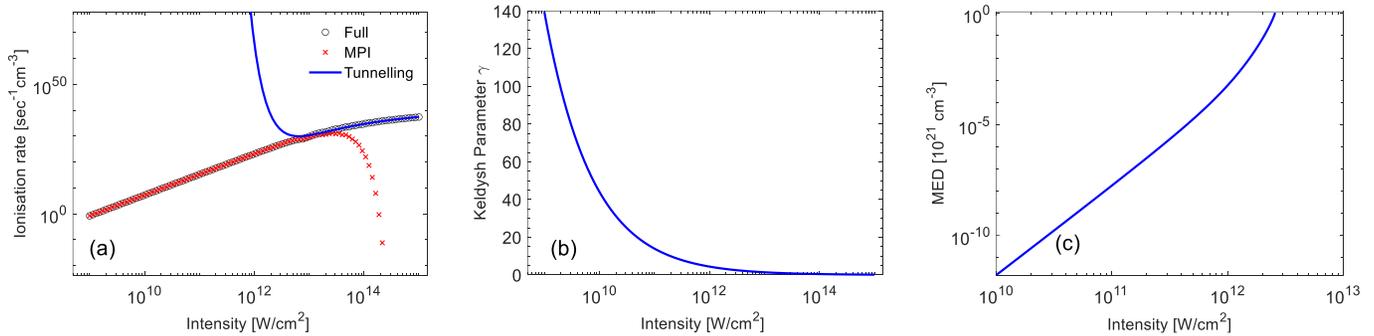

**Figure 7:** (a) Photo-ionisation rates and (b) Keldysh parameter and (c) MED as a function of the laser (peak) intensity. In (a) '*Full*' corresponds to the use of the photoionisation model without the employment of the approximating expressions for '*MPI*' (or '*Tunnelling*') processes [20, 22, 28, 33].

While the above analysis highlights the regimes in which different photo-ionization mechanisms dominate, offering insights into the fundamental physics of the interaction of the laser source with PC, it is also crucial to highlight the limitations of this methodology and recognize where the aforementioned modelling approach ceases to describe adequately the ultrafast phenomena. Furthermore, from a fundamental perspective, it is essential to understand which photo-ionization processes are favored when the irradiated material retains its original polymeric properties. Considering the behavior of polymers when exposed to high-intensity laser pulses—such as photo-chemical degradation, chain rearrangements, bond breaking, or polymerization—it is essential to identify the laser conditions that could trigger these complex transformations. Although several criteria can be associated with the onset of such changes, a condition that has been used in the previous studies (i.e. to describe damage in dielectric materials [19, 20], although, a thermal criterion is also considered, as it is believed to offer a more accurate assessment, even in those materials [18, 27], see Supplementary Material) is based on the evaluation of the intensity threshold at which a critical density of excited carriers is reached. This occurs when the simulated $N_e$ exceeds the critical value, defined as $N_e^{cr}$ (i.e. $N_e^{cr} \equiv 4\pi^2 c^2 m_e \varepsilon_0/(\lambda_L^2 e^2)$), the free electron density at which the plasma oscillation frequency is equal to the laser frequency ($N_e^{cr}$ =$1.1 \times 10^{21}$ cm⁻³, at $\lambda$=1026 nm). By calculating the maximum electron density (MED) across a range of intensities (Figure 7c), theoretical predictions indicate that the carrier density surpasses $N_e^{cr}$ for $I_{crit}$=$2.55 \times 10^{12}$ W/cm² (or equivalently, $F$=482 mJ/cm²). For intensities below $I_{crit}$, ionization of the material is predominantly conducted via multiphoton processes (Figure 7a).

## 5. Conclusions

This work represents a proof-of-concept to elucidate the photoionization process of polymeric materials excited by femtosecond pulses, using a theoretical model and femtosecond TAS experimental data. Key parameters of electron dynamics, including relaxation and recombination times, as well as predictions for the existence of STE states have, also, been evaluated while the employed TAS technique appears to provide significant insight into the carrier dynamics. Considering the complex behavior of polymers under exposure to ultrashort laser pulses, it is evident that a more comprehensive understanding of their response to laser excitation requires a deeper theoretical/experimental analysis. Moreover, a thorough investigation of the material's transformation upon exposure to high intensity pulses is essential for accurately modeling the patterning process. Nevertheless, the results obtained in the present work are expected to support future efforts aimed at elucidating how intense ultrashort laser pulses interact with polymeric materials which is crucial for optimizing the manufacturing processes of these materials for potential applications.

**CRediT authorship contribution statement**

**George D. Tsibidis**: conceptualization of the work (equal); physical modelling, simulations and interpretation of the results; writing-original draft (lead); **Matina Vlahou**: conceptualization of the work (equal); experimental protocol (equal); data acquisition (equal); TAS data processing and analysis (lead); writing-original draft (equal); **Stella Maragkaki**: experimental protocol (equal); data acquisition (equal); writing-original draft (equal); **Ioannis Konidakis**: experimental protocol (equal); data acquisition (equal); writing-original draft (equal); TAS experiment supervision; **Emmanuel Stratakis**: funding; writing-original draft (equal).

**Declaration of Competing Interest**

The authors declare that they have no known competing financial interests or personal relationships that could have appeared to influence the work reported in this paper.

**Data availability**

Data will be made available on request.

**Appendix A. Supplementary data**

Supplementary material related to this article can be found online at https://doi.org/......

**Acknowledgments**

This work has partially funded from Horizon Europe, the European Union's Framework Programme for Research and Innovation, under Grant Agreement No. 101057457 (*METAMORPHA*). G.D.T, also, acknowledges financial support from COST Action *PhoBioS* (supported by COST-European Cooperation in Science and Technology).